\documentclass[prb,aps,twocolumn,showpacs,home,floats]{revtex4}

\usepackage{graphics}
\begin{document}

\title{  {\rm\small\hfill (submitted to Phys. Rev. B})\\
Composition and structure of the RuO${}_2$(110) surface in an
O$_2$ and CO environment: implications for the catalytic formation of CO$_2$}

\author{Karsten Reuter and Matthias Scheffler}

\affiliation{Fritz-Haber-Institut der Max-Planck-Gesellschaft, Faradayweg
4-6, D-14195 Berlin-Dahlem, Germany}

\date{Received 27 January 2003}

\begin{abstract}
The phase diagram of surface structures for the model
catalyst RuO${}_2$(110) in contact with a gas environment of O$_2$ 
and CO is calculated by density-functional theory and atomistic 
thermodynamics. Adsorption of the reactants is found to depend crucially
on temperature and partial pressures in the gas phase. Assuming that a
catalyst surface under steady-state operation conditions is close to
a constrained thermodynamic equilibrium, we are able to rationalize a number
of experimental findings on the CO oxidation over RuO${}_2$(110).
We also calculated reaction pathways and energy barriers. Based on the 
various results the importance of phase coexistence conditions is 
emphasized as these will lead to an enhanced dynamics at the catalyst
surface. Such conditions may actuate an additional, kinetically controlled
reaction mechanism on RuO${}_2$(110).
\end{abstract}



\maketitle

\section{Introduction} 

A prerequisite for analyzing and understanding the electronic properties
and the function of surfaces, is the detailed knowledge of the atomic 
structure, i.e., the surface composition and geometry. In this respect 
the experimental techniques of ultra-high vacuum (UHV) {\em Surface
Science} have significantly helped to build our current 
understanding,\cite{masel96} and sometimes these UHV results
can be related to high-pressure applications, like catalysis or
corrosion. Often, however, such extrapolation is not straightforward and may
even be impossible.\cite{stampfl02} This is due to the fact, that some 
surface structures that can be stabilized in UHV  - sometimes by
sophisticated preparation, annealing, and post-dosing procedures - may not 
exist at high temperature, $T$, and high pressure, $p$. And structures that 
exist at high pressure may be hard or even impossible to identify or stabilize 
under UHV conditions. Obviously, knowing what structures dominate the 
high-pressure application one is interested in, is crucial for finding
a way to prepare the same or a similar situation under UHV conditions, 
in order to then perform controlled, atomistic experiments. Thus, the 
determination of a ($T, p$) phase diagram, covering the surface phases 
from UHV to realistic conditions appears to be critical for a meaningful
and safe bridging of the pressure gap, enabling a {\em Surface Science} 
study relevant to the high-pressure problem.

The concept of {\em first-principles atomistic thermodynamics} 
\cite{weinert86,scheffler88,kaxiras87,qian88} enables us to calculate
such surface phase diagrams, and for metal oxides this approach
has recently proven to be most valuable (see for example Refs.
\onlinecite{wang98,wang00,reuter02a,reuter03}). Using density-functional theory,
one calculates the free energies of all plausible surface 
compositions and geometries, in order to identify the lowest-energy
structure for a given condition of the thermodynamic reservoirs for the
atoms and electrons. In Ref. \onlinecite{reuter02a} we described the 
approach in detail for a one-component gas phase, and we now extend it 
to a multi-component environment also employing the concept of a 
``constrained thermodynamic equilibrium''; at the end we will also
analyze (and emphasize) the role of kinetic effects. A short paper on this
study already appeared.\cite{reuter03} 

We believe that the below discussion is relevant to oxidation catalysis at a 
ruthenium catalyst, but the applicability of the methodology is much wider. 
From the calculated surface phase diagram, a number of conclusions on the 
experimentally reported high efficiency of this oxide surface
\cite{cant78,peden86,boettcher97,boettcher99,over00,kim01b,fan01,wang01,wang02}
can already be drawn. Most notably, we emphasize that gas phase conditions,
that correspond to a coexistence of different surface phases, may lead to 
an enhanced dynamics and be particularly important for catalytic applications.
For the specific example of the RuO${}_2$(110) surface we find that such
phase coexistence conditions at high partial pressures enable a reaction 
mechanism, that does not play a role otherwise. Interestingly, the energy 
barrier is found to be even lower than that of other mechanisms considered 
so far.\cite{over00,liu01}

\section{Theory}

Our approach connects density-functional theory (DFT) total-energy
calculations and {\em atomistic thermodynamics}.
\cite{weinert86,scheffler88,kaxiras87,qian88}
In a preceding publication the procedure was described for an oxide 
surface in equilibrium with a one-component gas phase 
\cite{reuter02a}. We will therefore recapitulate it here only
briefly, concentrating now on the extension to a multi-component gas
phase and a ``constrained thermodynamic equilibrium''.
For clarity we will stick to the specific example of a 
RuO${}_2$(110) surface in contact with a gas phase environment formed of 
O$_2$ and CO. The generalization to other surfaces in contact with
arbitrary multi-component gas or even liquid phases is straightforward.

\subsection{Surface free energy} 

For a surface in equilibrium with atomic reservoirs (defined by a gas or 
liquid phase environment, or a macroscopic bulk phase), the most relevant 
structures are characterized by a low surface free energy, which is defined
as
\begin{equation}
\gamma(T,\{p_i\}) = \frac{1}{A} \left[ G \;-\;
\sum_i N_i \mu_i(T, p_i) \right].
\label{eq1}
\end{equation}

\noindent
Here, $G$ is the Gibbs free energy of the solid with the surface we like
to study. If a slab is used, there are two surfaces (the top and the bottom
side), both of which are then, of course, to be considered in the (total)
surface area, $A$. $\mu_i(T,p_i)$ is the chemical potential of the
species of $i$th type (here: $i =$ Ru, O$_2$, CO), and $N_i$ is the number
of atoms (or molecules) of the $i$th species in the considered solid. 
$T$ and $\{p_i\}$ are the temperature and the partial pressures of the 
various species. Considering the presence of two independent reservoirs of
O$_2$ and CO implies that O$_2$ and CO molecules, though their mix 
forms the environment, are not in equilibrium with each other:
From energy considerations alone, CO${}_2$ would result as the
most stable gas phase molecule for almost all temperature and pressure
conditions, {\em if} the environment were able to attain thermodynamic 
equilibrium with itself. However, the large free energy barrier for the 
gas phase reaction CO + 1/2 O${}_2 \rightarrow$ CO${}_2$ prevents that 
this reaction plays any role on the time scales of interest. 
Thus, our treatment ignores (because of good reasons) CO$_2$ formation 
in the gas phase, and only at the end of our study (sections III D-F) we will
consider that such a reaction may take place between species that are 
adsorbed on the surface. As a consequence, the oxygen and CO chemical 
potentials in this ``constrained thermodynamic equilibrium'' situation
are given by the expressions discussed in the next paragraph.

With respect to the Ru chemical potential we note the other constraint
of the present study, namely the presence of macroscopic quantities
of bulk RuO$_2$. If the temperature is not too low, this RuO$_2$ material
is in equilibrium with the O$_2$ environment which implies
\begin{equation}
\mu_{\rm Ru} \;+\; \mu_{{\rm O}{_2}} \;=\; g^{\rm bulk}_{\rm RuO_2}, 
\label{eq2}
\end{equation}
where $g^{\rm bulk}_{\rm RuO_2}(T,p_{\rm RuO_2})$ is the Gibbs free
energy of the bulk oxide (per formula unit). Inserting this into eq.
(\ref{eq1}) to eliminate $\mu_{\rm Ru}$ and rewriting it for a slab
calculation with two equivalent surfaces leads to a surface free 
energy, which is now a mere function of the chemical potentials 
determined by the equilibrium with the gas phase reservoirs, namely 
$\mu_{\rm CO}$ and $\mu_{\rm O_2}$,
\begin{eqnarray}
\label{eq3}
\lefteqn{\gamma(T,p_{{\rm O}{_2}}, p_{\rm CO}) \;=\; \frac{1}{A}}&& \\ \nonumber
&& \bigg[ \left( \; G^{\rm surf}_{\rm slab}(T,\{p_i\},\{N_i\}) -
N_{\rm Ru} g^{\rm bulk}_{\rm RuO_2}(T,p_{\rm RuO_2}) \; \right) + \\ \nonumber 
&& + \left( 2 N_{\rm Ru} - N_{\rm O} \right) \mu_{\rm O}(T,p_{{\rm O}}{_2}) \;-\;
N_{\rm CO} \mu_{\rm CO}(T,p_{\rm CO}) \; \bigg] .
\end{eqnarray}

\noindent
Here we also used that $ N_{\rm O} = 2 N_{{\rm O}{_2}}$ and
$\mu_{\rm O} = (1/2) \mu_{{\rm O}{_2}}$. The Gibbs free energies of the 
slab and of the RuO$_2$ bulk can be calculated by density-function theory, 
evaluating the total energies and the vibrational spectra.\cite{reuter02a} 
As apparent from eq. (\ref{eq3}), $\gamma(T,p_{{\rm O}{_2}}, p_{\rm CO})$
depends only on the {\em difference} of the Gibbs free energies of the 
bulk and the slab results. In our previous work we could show that for
RuO${}_2$ the vibrational energy and entropy contributions to this 
{\em difference} of Gibbs free energies cancel to a large extent\cite{reuter02a},
and we will therefore replace the slab and bulk Gibbs free energies by
the corresponding total energies. We note, however, that for other systems 
this procedure can cause a noticeable error. Furthermore, we note (details 
are discussed in the appendix) that some surface phases may be disordered 
which gives rise to configurational entropy. On the energy scale relevant to 
the present study, this configurational-entropy contribution is negligible 
(cf. the appendix).

\subsection{Chemical potentials}

The chemical potentials of O and CO, that enter eq. (\ref{eq3}), are 
determined by the condition of thermodynamic equilibrium with the
surrounding gas phase reservoirs. Thus, their temperature and pressure 
dependence is
\begin{equation}
\mu_{\rm O}(T,p_{\rm O_2}) = 
\frac{1}{2}\left[ E^{\rm total}_{\rm O_2}
+ \tilde{\mu}_{\rm O_2}(T,p^{0}) 
+ k_{\rm B} T ln \left( \frac{p_{\rm O_2}}{p^{0}} \right) \right],
\label{eq4}
\end{equation}
and
\begin{equation}
\mu_{\rm CO}(T,p_{\rm CO}) =
E^{\rm total}_{\rm CO}
 + \tilde{\mu}_{\rm CO}(T,p^{0})
+ k_{\rm B} T ln \left( \frac{p_{\rm CO}}{p^{0}} \right).
\label{eq5}
\end{equation}
The temperature dependence of $\tilde{\mu}_{{\rm O}{_2}}(T,p^{0})$
and $\tilde{\mu}_{\rm CO}(T,p^{0})$ includes the contributions from
vibrations and rotations of the molecules, as well as the ideal gas
entropy at 1 atmosphere. It can be calculated, but in this paper we 
will simply use the experimental values from thermodynamic 
tables\cite{reuter02a,JANAF}, given in Table \ref{tableI}.

The $T=0$ K values of the O$_2$ and CO chemical potentials, 
$E^{\rm total}_{\rm O_2}$ and $E^{\rm total}_{\rm CO}$, are the total 
energies of the isolated molecules (including zero point vibrations),
and for these we will use our DFT results. We note that state-of-the-art 
DFT total energies of the O$_2$ molecule are subject to a noticeable error.
Again, for the present study this error is not crucial and can partially
be circumvented \cite{reuter02a}, however for other systems it maybe 
important to treat $E^{\rm total}_{\rm O_2}$ differently (see, for example, 
the case of silver oxide discussed in Ref. \onlinecite{Li03}).

\begin{table}
\caption{\label{tableI}
$\tilde{\mu}_{{\rm O}{_2}}(T,p^{0})$ and $\tilde{\mu}_{\rm CO}(T,p^{0})$ 
in the temperature range of interest to our study. The employed entropy 
and enthalpy changes are taken from the JANAF thermochemical tables at 
$p^{0} = 1 \; {\rm atm}$.\cite{JANAF}}
 
\begin{tabular}{rrr | rrr}
$T$   & $\tilde{\mu}_{\rm O_2}(T,p^{0})$ & $\tilde{\mu}_{\rm CO}(T,p^{0})$ & 
$T$   & $\tilde{\mu}_{\rm O_2}(T,p^{0})$ & $\tilde{\mu}_{\rm CO}(T,p^{0})$ \\ \hline
100 K & -0.16 eV &  -0.14 eV  & 600 K  & -1.22 eV & -1.18 eV \\
200 K & -0.34 eV &  -0.33 eV  & 700 K  & -1.46 eV & -1.40 eV \\
300 K & -0.54 eV &  -0.53 eV  & 800 K  & -1.70 eV & -1.64 eV \\
400 K & -0.76 eV &  -0.73 eV  & 900 K  & -1.98 eV & -1.88 eV \\
500 K & -1.00 eV &  -0.95 eV  &1000 K  & -2.20 eV & -2.12 eV \\
\end{tabular}
\end{table}

In the following we will present the resulting surface energies as a 
function of the chemical potentials, i.e. in $(\mu_{\rm O},
\mu_{\rm CO})$-space. Equations (\ref{eq4}) and (\ref{eq5}) describe how 
these chemical potentials can be converted into pressure scales at any
specific temperature. We will exemplify this below by showing pressure
scales at $T=300$\,K and $T=600$\,K in order to  elucidate
the physical meaning behind the obtained results.

\subsection{DFT computations}

\begin{figure}
\scalebox{0.75}{\includegraphics{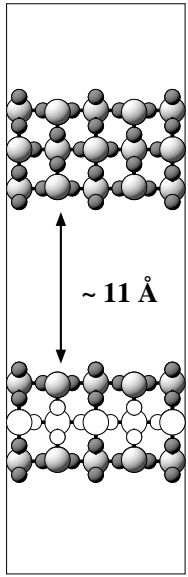}}
\caption{Sideview of the chosen supercell setup (Ru = light,
large spheres, O = dark, medium spheres). Each slab consists
of three O-(RuO)-O trilayers, and the atoms of one such trilayer
have been whitened in the lower slab. Consecutive slabs are
separated by a vacuum region of $\approx 11$\,{\AA}.}
\label{fig0}
\end{figure}

The total energies entering eq. (\ref{eq3}) are obtained by 
DFT calculations using the full-potential linear augmented plane wave
method (FP-LAPW) \cite{blaha99,kohler96,petersen00} together with the
generalized gradient approximation (GGA) for the exchange-correlation
functional \cite{perdew96}. The RuO${}_2$(110) surface is modeled
in a supercell geometry, employing a symmetric slab consisting of
three rutile O-(RuO)-O trilayers as shown in Fig. \ref{fig0}. All
atomic positions within the outermost trilayer were fully relaxed. 
A vacuum region of $\approx$ 11 {\AA} ensures the decoupling of the
surfaces of consecutive slabs as described previously.\cite{reuter02a}

The FP-LAPW basis set parameters are: $R_{\rm{MT}}^{\rm{Ru}}=$1.8
bohr, $R_{\rm{MT}}^{\rm{O}}=$1.1 bohr, $R_{\rm{MT}}^{\rm{C}}=$1.0 bohr,
wave function expansion inside the muffin tins up to 
$l_{\rm{max}}^{\rm{wf}} = 12$, potential expansion up to 
$l_{\rm{max}}^{\rm{pot}} = 4$. For the RuO${}_2$(110) slabs the 
$(1 \times 1)$ Brillouin zone (BZ) integration was performed using a $(5
\times 10 \times 1)$ Monkhorst-Pack grid with 50 (15) {\bf k}-points in
the full (irreducible) part of the BZ. The energy cutoff for the plane
wave representation in the interstitial region between the muffin tin 
spheres was $E^{\rm max}_{\rm wf}$ = 20 Ry for the wave functions and 
$E^{\rm max}_{\rm pot}$ = 169 Ry for the potential. With one notable 
exception this is exactly the same basis set as used in our preceding 
work on RuO${}_2$(110) in contact with a pure O environment 
\cite{reuter02a}. The calculated short CO bondlength of 1.15
{\AA} (exp: 1.13 {\AA}) forced us to reduce the oxygen muffin-tin
radius from 1.3\,bohr to 1.1\,bohr, so that we had to increase 
$E^{\rm max}_{\rm wf}$ from 17 Ry to 20 Ry to achieve the same high level
of convergence as detailed before.

With this basis set, the computed binding energies for the free 
O${}_2$ ($E^{\rm bind}_{\rm O_2} = -6.07$\,eV), for the 
CO ($E^{\rm bind}_{\rm CO} = -11.42$\,eV), 
and for the CO${}_2$ ($E^{\rm bind}_{\rm CO_2} = -17.62$ eV)
molecule are in very good agreement with previously reported DFT values, 
while still showing the known overbinding with respect to experiment 
\cite{ganduglia99,eichler99,CRC95}. Note, that the above values contain
zero-point vibrations that were estimated as 0.09 eV, 0.13 eV and
0.30 eV respectively. By comparison with test calculations performed 
at $E^{\rm max}_{\rm wf} =$\,24 Ry, we conclude that the absolute binding
energies of O and CO at the RuO${}_2$(110) surface are converged within
0.15\,eV/atom, which translates into surface free energy variations below
10 meV/{\AA}${}^2$. For the {\em differences} between the computed
$\gamma(T,\{p_i\})$ of different surface phases, which are the quantities
actually entering into the construction of the phase diagram presented here,
the error cancellation is even better. As a consequence, the numerical 
uncertainty of results discussed below is $\pm 5$\,meV/{\AA}${}^2$
(with respect to the basis set and the supercell approach). While this
error bar does not affect any of the physical conclusions drawn, we note 
that it does not include the errors introduced by the more basic deficiency 
of density-functional theory, namely the approximate nature of the employed 
exchange-correlation functional, the effect of which will be discussed
in section III C.

\section{Results}

\subsection{Stability range of RuO${}_2$}

As we are interested in the adsorption of reactants on a stable 
RuO$_2$ substrate, we first analyze the stability range of rutile
RuO${}_2$ bulk. In our previous work \cite{reuter02a} we had already 
shown that in a pure oxygen environment, the $\mu_{\rm O}$-variations 
may be restricted to a finite range: Below the so-called ``O-poor limit'' 
the oxide will decompose into Ru metal and oxygen, i.e., the oxide is 
only stable if
\begin{equation}
g^{\rm bulk}_{\rm RuO_2} \; <  \; g^{\rm bulk}_{\rm Ru} + 2 \mu_{\rm O},
\end{equation}
i.e., if
\begin{equation}
\Delta \mu_{\rm O} > \frac{1}{2}[g^{\rm bulk}_{\rm RuO_2} 
- g^{\rm bulk}_{\rm Ru}
- E^{\rm total}_{\rm O_2}].
\label{O2-condition}
\end{equation}
Here $\Delta \mu_{\rm O}$ is defined as $\mu_{\rm O} - (1/2) 
E^{\rm total}_{\rm O_2}$. For $T=0$ K the right hand side equals half
of the low temperature limit of the heat of formation, 
$H_f(T=0\,{\rm K}, p=0)$, for which our 
DFT calculations give $H_f(T=0\,{\rm K}, p=0) = -3.4$ eV (the
experimental result is $-3.19$ eV \cite{CRC95}). Thus, despite the
mentioned error in $E^{\rm total}_{\rm O_2}$, for $H_f$ there is a 
fortuitous error cancellation. We note in passing that for the bulk 
phases the $T$ and $p$ dependence of the Gibbs free energies is small. 
Thus, replacing the right side of eq. (\ref{O2-condition}) by 
$(1/2)H_f(T=0\,{\rm K})$ is a good approximation.

The ``O-rich limit'' refers to conditions where oxygen will condense, and
this gives the other restriction:
\begin{equation}
\Delta\mu_{\rm O} < 0.
\label{O2-rich}
\end{equation}
Combining the above two equations gives the range of $\Delta \mu_{\rm O}$ 
we will consider in the later discussion as
\begin{equation}
\frac{1}{2} H_f(T=0{\rm K},\; p=0)
\;<\; \Delta\mu_{\rm O}(T, p_{{\rm O}_2})
\;<\; 0.
\label{eq6}
\end{equation}

RuO2 can also be destroyed (reduced) by carbon monoxide. In a pure CO 
environment the stability condition for the oxide is 
\begin{equation}
g^{\rm bulk}_{\rm RuO_2} + 2\mu_{\rm CO} \; <  \; g^{\rm bulk}_{\rm Ru} +
2E^{\rm total}_{\rm CO_2}.
\label{E-CO2}
\end{equation}
This can be written as
\begin{equation}
\Delta\mu_{\rm CO}  < - \frac{1}{2}[g^{\rm bulk}_{\rm RuO_2}
- g^{\rm bulk}_{\rm Ru} - E^{\rm total}_{\rm O_2}]
- \frac{1}{2}E^{\rm bind}_{\rm O_2} - E^{\rm bind}_{\rm CO}
+ E^{\rm bind}_{\rm CO_2}.
\label{CO-condition}
\end{equation}
Here $\Delta \mu_{\rm CO}$ is defined as $\mu_{\rm CO} -
E^{\rm total}_{\rm CO}$. For the right hand side of eq. 
(\ref{CO-condition}) (at $T=0$ K) our DFT result is $-1.5$ eV 
(the experimental result is $-1.33$ eV \cite{JANAF}).

Let us now consider that O$_2$ and CO are both present in the gas 
phase. However, as mentioned earlier, we will consider the oxide 
(and its surface) in what we call a ``constrained equilibrium'', 
i.e. we consider that RuO$_2$ is in thermodynamic equilibrium with 
O$_2$ and with CO individually, while the O${}_2$ and CO in the gas
phase are not equilibrated. Combining eqs. (\ref{O2-condition}) and 
(\ref{CO-condition}) defines a line in $(\Delta \mu_{\rm O}, 
\Delta \mu_{\rm CO})$-space:
\begin{equation}
\Delta \mu_{\rm O} - \Delta \mu_{\rm CO} =  C,
\label{instability}
\end{equation}
with $C= - 0.2$ eV (theory) and $- 0.26$ eV (experiment)\cite{JANAF}.
Above the line, i.e. for $\Delta\mu_{\rm CO} > \Delta\mu - C$,
RuO$_2$ decomposes into Ru and below the line the oxide is stable.
The situation close to the line will be discussed, also for other 
oxides, in a subsequent publication.

Let us add a comment on the above treatment of the CO$_2$ molecules.
As mentioned above, CO$_2$ is assumed to be not in thermodynamic 
equilibrium with any reservoir or any other species. A dynamical
treatment of the CO${}_2$ gas flow is outside the scope of this
paper, and as a first estimate we therefore use the single-molecule 
internal energy (total energy plus vibrations and rotations) in eq. 
(\ref{E-CO2}) and in the following.

Corresponding to the thus derived stability conditions we will only show
our computed surface free energies in the limited range $-1.68$ eV $< 
\Delta \mu_{\rm O}(T,p) <$ 0.0 eV and $-2.0$ eV $< 
\Delta \mu_{\rm CO}(T,p) < 0.0$ eV, and mark the instability line, eq. 
(\ref{instability}), by a white dotted line. Finally, we also note that 
under very reducing conditions (low $\Delta \mu_{\rm O}$, high 
$\Delta \mu_{\rm CO}$) CO will transfrom into graphite, which then 
determines the C chemical potential and which e.g. occurs for 
$\Delta \mu_{\rm CO} = 0.0$\,eV for any $\Delta \mu_{\rm O} < -1.2$\,eV. 
The corresponding region will be marked by extra hatching in our graphs, 
although it corresponds to gas phase conditions where RuO${}_2$ is 
already only metastable with respect to the aforedescribed CO-induced 
decomposition.

\subsection{Adsorption of O and CO on RuO${}_2$(110)}

\begin{figure}
\scalebox{0.5}{\includegraphics{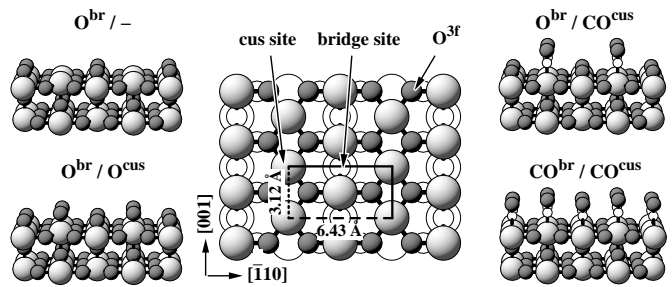}}
\caption{Top view of the RuO${}_2$(110) surface explaining the
location of the two prominent adsorption sites (bridge and cus),
which continue the bulk-stacking sequence (central panel).
Also shown are perspective views of the two most stable
terminations involving only O (left panel) and involving O and
CO (right panel). Ru = light, large spheres, O = dark, medium
spheres, C = white, small spheres. Atoms lying in deeper layers
have been whitened in the top view for clarity.}
\label{fig1}
\end{figure}

In the rutile RuO${}_2$ bulk structure, every Ru atom has
six oxygen neighbors and every oxygen atom three metal 
neighbors \cite{reuter02a,sorantin92}. The RuO${}_2$(110) 
surface structure is conveniently explained on the basis of the 
mixed (RuO) plane termination exhibiting threefold coordinated 
O${}^{\rm 3f}$ lattice oxygen atoms, cf. the middle panel in Fig.
\ref{fig1}. The computed lattice constants for the rectangular
$(1 \times 1)$ surface unit-cell are (6.43\,{\AA} $\times$
3.12\,{\AA}), which compare well with the experimentally
reported values of (6.35\,{\AA} $\times$ 3.11\,{\AA}) 
\cite{sorantin92,atanasoska88}. The surface exhibits two
distinct adsorption sites, namely the so-called coordinatively
unsaturated (cus) site atop of fivefold coordinated Ru atoms
(Ru${}^{\rm cus}$), as well as the bridge site between two
fourfold coordinated Ru atoms (Ru${}^{\rm br}$), cf. Fig. \ref{fig1}.
The bulk stacking of the oxide would be continued by oxygen
atoms first occupying all bridge sites (O${}^{\rm br}$), this
way yielding the stoichiometric surface termination (upper left 
panel of Fig. \ref{fig1}) that also arises naturally when cutting
the bulk between two of the subsequent O-(RuO)-O trilayers shown
in Fig. \ref{fig0}. Occupying also all cus sites (O${}^{\rm cus}$)
then leads to the third possible $(1 \times 1)$ termination of a
rutile (110) surface shown in the lower left panel of Fig. 
\ref{fig1}.\cite{reuter02a}

Adsorption of either O or CO is in principle also conceivable
at other sites on the surface. However, extensive calculations
with O and CO in such low-symmetry sites showed significantly 
lower binding energies compared to adsorption at the cus and 
bridge sites, which is why we will restrict our discussion to 
these two prominent sites in the following. We also performed a
large number of calculations in larger supercells up to 
$(1 \times 4)$ and $(2 \times 2)$ periodicity to assess the
importance of lateral interactions between the adsorbates. The 
obtained binding energies for both adsorbates varied by less 
than 150\,meV \cite{reuter02c}, reflecting rather weak lateral 
interactions beyond the direct first neighbor cus-br interaction 
(implicitly contained within our DFT-computed total energies of 
$(1 \times 1)$ phases). Correspondingly, only the latter phases
will be considered in the following.

\begin{table}
\caption{\label{tableII}
Calculated binding energies in $(1 \times 1)$ phases on 
RuO${}_2$(110). The values are given with respect to the
free O${}_2$ and CO molecule respectively, zero-point
vibrations not included.}
\begin{tabular}{l | l | l}
Species          & Phase                             & Binding Energy \\ \hline
O${}^{\rm br}$   & O${}^{\rm br}$/--                 & -2.44\,eV/atom  \\
O${}^{\rm br}$   & O${}^{\rm br}$/O${}^{\rm cus}$    & -2.23\,eV/atom  \\
O${}^{\rm br}$   & O${}^{\rm br}$/CO${}^{\rm cus}$   & -2.37\,eV/atom  \\
O${}^{\rm cus}$  & O${}^{\rm br}$/O${}^{\rm cus}$    & -0.99\,eV/atom  \\
O${}^{\rm 3f}$   & O${}^{\rm br}$/--                 & -3.59\,eV/atom  \\
CO${}^{\rm cus}$ & O${}^{\rm br}$/CO${}^{\rm cus}$   & -1.38\,eV/atom  \\
CO${}^{\rm cus}$ & CO${}^{\rm br}$/CO${}^{\rm cus}$  & -1.39\,eV/atom  \\
CO${}^{\rm br}$  & CO${}^{\rm br}$/CO${}^{\rm cus}$  & -1.58\,eV/atom  \\
\end{tabular}
\end{table}

Depending if the bridge and cus site within a $(1 \times 1)$ cell
is either occupied by O or CO or empty, nine different adsorption
geometries result. For these we introduce a short-hand notation
indicating first the occupancy of the bridge and then
of the cus site, e.g. O${}^{\rm br}$/-- for O adsorption
at the bridge site, the cus site being empty. Having computed
the total energies of all nine combinations, we only find
the four depicted in perspective views in the side panels in
Fig. \ref{fig1} to be relevant in the context of a high
pressure gas phase and will consequently dedicate ourselves
to them for clarity. 

From our calculations we obtain a number of $T = 0$\,K binding
energies, which are given in Table \ref{tableII}. Good agreement for
CO is obtained with other DFT calculations 
\cite{kim00,kim01,seitsonen01,liu01}. Also the value for 
O${}^{\rm br}$ agrees nicely with the value reported by Liu 
{\em et al.} \cite{liu01}. On the other hand, all oxygen binding 
energies are $\sim 0.8$\,eV stronger than those reported by 
Seitsonen and coworkers \cite{kim00,kim01,seitsonen01}. We
are unable to explain this discrepancy, but note that our 
O${}^{\rm cus}$ binding energy of about -1 eV (with respect to 
$1/2 E^{\rm total}_{\rm O_2}$) is consistent with the 
thermodesorption temperature of $T = 300-550$\,K reported for
this species.\cite{boettcher99}

From the numbers listed in Table \ref{tableII} we see that O
adsorption at bridge sites is in general significantly stronger 
than adsorption at the cus sites, which is comprehensible
considering the on-top, onefold bonding geometry at the cus
sites compared to the bridge, twofold bonding geometry at the 
bridge sites, cf. Fig. \ref{fig1}. Interestingly, CO does not follow such a
pronounced bond-order conservation trend and exhibits only
slight differences in bond strength at bridge and cus sites.
As has already been pointed out by Seitsonen {\em et al.}, this 
might be related to the specific CO$^{\rm br}$ adsorption geometry,
which in the here considered $(1 \times 1)$ phase is an asymmetric
bridge with the CO much stronger bound to only one of the two 
neighboring Ru$^{\rm br}$ atoms.\cite{seitsonen01}

\subsection{Surface phase diagram}

\begin{figure}
\scalebox{0.35}{\includegraphics{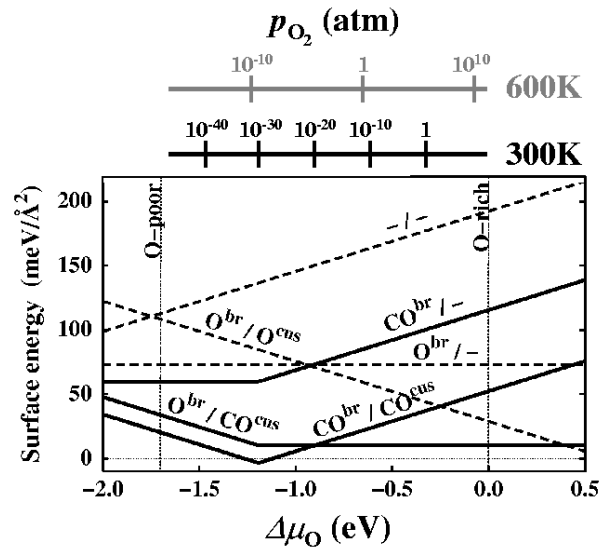}}
\caption{Surface free energies of the three only O involving
RuO${}_2$(110)-$(1 \times 1)$ terminations (dashed lines) and
of three O and CO involving surface geometries (solid lines).
Corresponding geometries are explained in the left and right
panels of Fig. 1, respectively. The dotted vertical lines
indicate the allowed range of oxygen chemical potential,
cf. eq. (\ref{eq6}), while $\Delta \mu_{\rm CO} = 0.0$\,eV,
corresponding to very CO-rich conditions. In the top $x$-axis, 
the dependence on $\Delta \mu_{\rm O}(T,p_{\rm O})$ has been cast 
into pressure scales at fixed temperatures of $T = 300$\,K
and $T = 600$\,K.}
\label{fig2}
\end{figure}

Before proceeding to set up the surface phase diagram by evaluating
eq. (\ref{eq3}) for any $(\mu_{\rm O},\mu_{\rm CO})$, we start by 
summarizing the previously reported effect of a just oxygen
containing environment \cite{reuter02a}. In this special case,
corresponding to $\Delta \mu_{\rm CO} \rightarrow - \infty$, we have
to consider the stability of the aforedescribed three possible
$(1 \times 1)$ oxide terminations --/--, O${}^{\rm br}$/-- and 
O${}^{\rm br}$/O${}^{\rm cus}$ as a function of the oxygen chemical
potential. From the computed surface free energies drawn as dashed
lines in Fig. \ref{fig2} we see that in the allowed range of 
$\Delta \mu_{\rm O}$ delimited by the two vertical dotted lines, cf. eq.
(\ref{eq6}), the mixed (RuO) plane termination --/-- shown in the 
topview in Fig. \ref{fig1} is never stable, so that oxygen atoms will 
always occupy at least all bridge sites, O${}^{\rm br}$/--, as long as
the surface is in equilibrium with the surrounding gas phase. Towards
higher $\Delta \mu_{\rm O}$ oxygen atoms will additionally occupy all cus
sites, leading to the polar O${}^{\rm br}$/O${}^{\rm cus}$ termination,
cf. Fig. \ref{fig2}. To give an impression of the corresponding
pressures required to stabilize this latter termination, we have also 
included in Fig. \ref{fig2} pressure scales as second $x$-axis for 
$T= 300$\,K and $T=600$\,K, the latter temperature corresponding to 
a typical annealing temperature frequently employed in experimental 
studies on this system
\cite{boettcher99,over00,kim01b,fan01,wang02,kim00,kim01,seitsonen01}.
From these scales we deduce that the stoichiometric O${}^{\rm br}$/-- 
termination, that was frequently prepared and studied in UHV studies
\cite{boettcher99,over00,kim01b,fan01,wang02,kim00,kim01,seitsonen01},
is primarily the end result of a high temperature anneal in UHV,
whereas at atmospheric O${}_2$ pressures the surface exhibits 
additionally oxygen atoms at the cus sites.

Extending our analysis now to the two-component (O${}_2$,CO) environment,
we fix the CO chemical potential first to $\Delta \mu_{\rm CO} = 0.0$\,eV,
corresponding to very CO-rich conditions. This allows us not only to
conveniently draw the surface free energies of three most stable O and CO
containing geometries into Fig. \ref{fig2} (solid lines), but
gives also the low $\gamma(T,\{p_i\})$-limit for these phases. As explained 
in Section IIIA, in this CO-rich limit the C chemical potential is 
determined by the equilibrium with graphite below $\Delta \mu_{\rm O} = 
- 1.2$\,eV and with CO above, which is why the corresponding three lines 
in Fig. \ref{fig2} exhibit a kink at this value. Even in the shown CO-rich 
limit, the CO$^{\rm br}$/-- phase due to CO adsorption at only the 
energetically more favorable bridge sites, cf. Table \ref{tableII}, is 
barely more stable than the hitherto considered pure O-terminations. This 
changes, if CO is additionally present also at the cus sites, as we find 
the completely CO covered CO$^{\rm br}$/CO$^{\rm cus}$ surface to have a 
low surface free energy at least at low O chemical potential. Towards 
higher $\Delta \mu_{\rm O}$ the third geometry exhibiting O atoms at the 
bridge sites and CO at the cus sites, O$^{\rm br}$/CO$^{\rm cus}$, becomes 
finally even more stable, cf. Fig. \ref{fig2}.

\begin{figure}
\scalebox{0.33}{\includegraphics{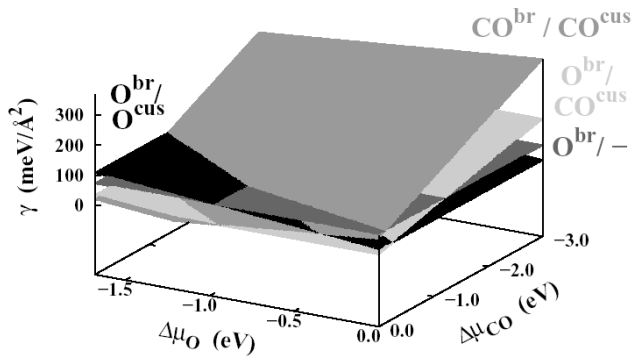}}
\caption{Surface free energies of the four stable geometries
shown in Fig. 1, as a function of ($\Delta \mu_{\rm O}, \Delta \mu_{\rm CO}$).}
\label{fig3}
\end{figure}

Allowing (in addition to $\Delta \mu_{\rm O}$) variations of 
$\Delta \mu_{\rm CO}$ away from the CO-rich limit, the two-dimensional 
graph in Fig. \ref{fig2} is extended into the three-dimensional one shown 
in Fig. \ref{fig3}, in which $\gamma(T,\{p_i\})$ is given as a function 
of $(\Delta \mu_{\rm O},\Delta \mu_{\rm CO})$. This implies obviously, 
that the contents of Fig. \ref{fig2} are contained as a special case in 
Fig. \ref{fig3}, in fact as the surface free energy dependence along the 
front $x$-axis of the latter figure. With decreasing CO chemical potential,
i.e. lower CO content in the gas phase, the CO involving surface phases 
become increasingly less favorable, whereas the surface free energies of
the two pure O terminations do of course not depend on $\Delta \mu_{\rm CO}$.
It is also worth pointing out that according to eq. (\ref{eq3}) the
slope with respect to $\Delta \mu_{\rm CO}$ of the plane representing the
CO$^{\rm br}$/CO$^{\rm cus}$ phase in Fig. \ref{fig3} is twice that
of the plane of the O$^{\rm br}$/CO$^{\rm cus}$ phase, given that the prior
structure contains double the amount of CO at the surface.
Although $\Delta \mu_{\rm CO}$ could in principle be varied down to 
$- \infty$, we only show in Fig. \ref{fig3} the range to $-3.0$\,eV, as 
already at this value the pure O terminations have become most stable for
any $\Delta \mu_{\rm O}$, indicating that the CO content in the gas phase
has become so low, that no CO can be stabilized at the surface anymore.

\begin{figure}
\scalebox{0.4}{\includegraphics{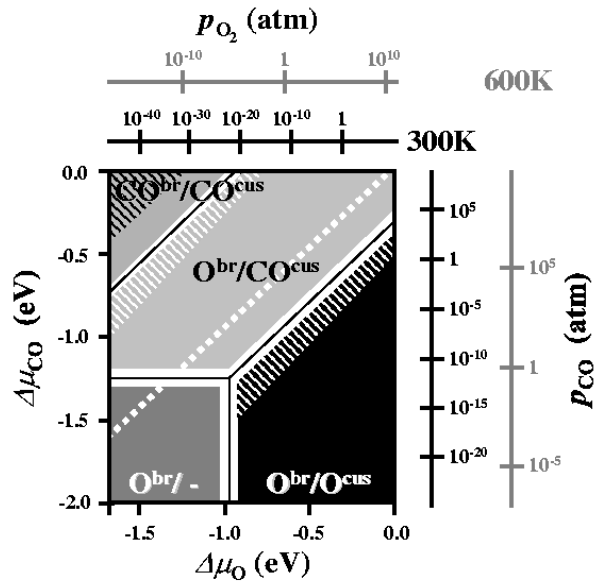}}
\caption{Surface phase diagram of stable and metastable structures of
RuO${}_2$(110) in $(\mu_{\rm O}, \mu_{\rm CO})$-space. 
The additional axes at the top/right give the corresponding
pressure scales at $T = 300$\,K and $600$\,K. The white regions 
close to the boundaries between different stable phases mark phase
coexistence conditions, where at least 10\% of the respective
other phase is also present at the surface at $T=300$\,K. Regions 
marked by white hatching are particularly strongly affected by kinetics
(see text). The white dotted line marks the stability limit of
bulk RuO${}_2$ with respect to CO-induced decomposition: In the
upper left part above this line, RuO${}_2$ is only metastable.}
\label{fig4}
\end{figure}

Although Fig. \ref{fig3} is instructive for understanding the extension
of the surface free energy dependence from a one-component to a 
two-component environment, the three-dimensional nature of the plot 
does not allow an easy access to the really important information 
contained in it, i.e. which phase has the lowest $\gamma(T,\{p_i\})$
and is thus most stable for a given gas phase characterized by 
$\Delta \mu_{\rm O}$ and $\Delta \mu_{\rm CO}$. This information is 
better obtained by only drawing the stability regions of the most 
stable phases, which corresponds to a projection of the lowest surface
free energies in Fig. \ref{fig3} onto the $(\Delta \mu_{\rm O},
\Delta \mu_{\rm CO})$-plane. The resulting {\em surface phase diagram} 
is drawn in Fig. \ref{fig4}. At the lowest CO chemical potentials shown, 
so little CO is present in the gas phase, that we simply recover the 
previously discussed surface structure dependence on $\mu_{\rm O}$ in 
a pure O${}_2$ atmosphere. In other words, the stoichiometric 
O${}^{\rm br}$/-- phase is stable at low O chemical potential, while 
higher O${}_2$ pressures stabilize additional oxygen at the surface, 
leading to the O${}^{\rm br}$/O${}^{\rm cus}$ phase. Increasing
the CO content in the gas phase, CO is first bound at the cus sites.
This is easier at low O chemical potentials, where these sites are
free, but is harder at higher $\Delta \mu_{\rm O}$ where also oxygen 
atoms compete for adsorption at the cus sites. As a result, towards 
higher $\Delta \mu_{\rm O}$ the O${}^{\rm br}$/CO${}^{\rm cus}$ phase 
becomes only more stable than the O${}^{\rm br}$/O${}^{\rm cus}$ phase
at progressively higher CO chemical potentials, cf. Fig. \ref{fig4}. 
Finally, under very reducing conditions (high $\Delta \mu_{\rm CO}$, 
low $\Delta \mu_{\rm O}$) CO is able to substitute O also at the 
bridge sites, yielding the completely CO covered 
CO${}^{\rm br}$/CO${}^{\rm cus}$ surface. Additionally shown in Fig.
\ref{fig4} is the line marking the instability of RuO${}_2$ with
respect to CO-assisted decomposition at high $\Delta \mu_{\rm CO}$ (white
dotted line). From the location of this line we see that the complete
stability region of the CO${}^{\rm br}$/CO${}^{\rm cus}$ phase 
(including the graphite-formation region hatched in black) and
large parts of the O${}^{\rm br}$/CO${}{\rm cus}$ phase correspond
already to metastable situations that will not prevail for long
under realistic conditions.

The surface phase diagram in Fig. \ref{fig4} summarizes the key results
of the present work. However, before proceeding to discuss the
physics contained in it, we like to discuss the accuracy of our theory.
As stated in section IIC, the numerical uncertainty in the surface free
energies due to the DFT basis set and the supercell approach is about
$\pm 5$ meV/{\AA}${}^2$. If we allow the four planes shown in Fig. \ref{fig3}
to shift by this value with respect to each other, the resulting
intersections, i.e. the phase boundaries, would shift within the white
regions drawn in Fig. \ref{fig4}. We have added additional axes
in Fig. \ref{fig4} indicating the pressure scales at $T=300$\,K and
600\,K. From these scales it becomes apparent, that at room temperature
our computational uncertainty may very well correspond to even a few
orders of magnitude in pressure. On the other hand, this is only so, 
because at $T = 300$\,K these one or two orders of magnitude in 
pressure sample only a vanishingly small part of phase space. In 
general, we believe that Fig. \ref{fig4} illustrates quite nicely
that our numerical inaccuracy does not affect the general structure of 
the obtained phase diagram at all - even if it would, it would just mean
that we would have to increase our basis set. 

\begin{figure}
\scalebox{0.4}{\includegraphics{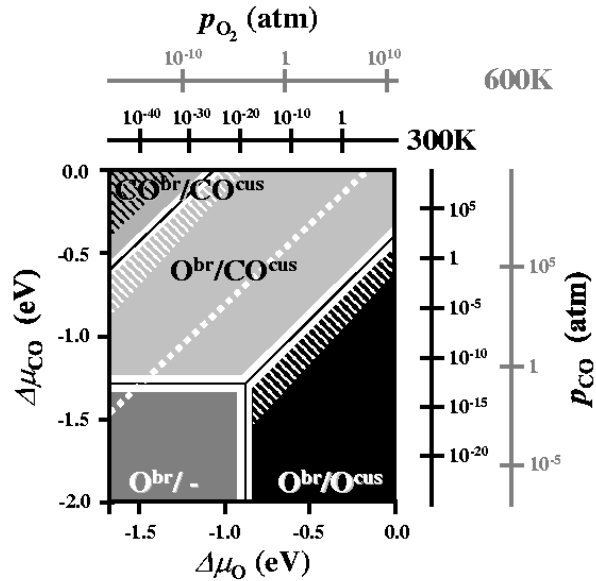}}
\caption{Same as Fig. 4, but now computed using the LDA
as exchange-correlation functional.} 
\label{fig5}
\end{figure}

This error analysis does not yet include the more basic deficiency of 
DFT, namely the approximate nature of the exchange-correlation functional. 
To this end, we have also computed the surface phase diagram not with a 
gradient-corrected functional, but with the local-density approximation
(LDA) \cite{perdew92}. The result (obtained by a complete geometry 
reoptimization of all structures using the LDA lattice constants) is 
shown in Fig. \ref{fig5} and is very similar to the 
GGA results (Fig. \ref{fig4}). It is worth pointing out that the absolute 
surface free energies for the different phases obtained within the LDA
are up to 20 meV/{\AA}${}^2$ higher than the GGA results. However, their
differences, which ultimately determine the positions of the phase 
boundaries, differ only little. Although this LDA and GGA comparison
does not provide a quantitative assessment of the remaining error that
is due to the approximate treatment of exchange and correlation,
the obtained small difference strongly suggests that the DFT-accuracy
for this phase diagram is rather high. In line with our previous
analysis \cite{reuter02a} we therefore conclude that the approximate
nature of the exchange-correlation functional may affect the exact
phase transition temperatures and pressure (with a typical uncertainty
of $\pm 100$\,K and 1-2 orders of magnitude in pressure), but
does not affect the overall structure of the obtained phase diagram.

\subsection{Surface phase diagram, kinetics and catalysis}

The main outcome of the preceding section is the surface phase diagram
of RuO${}_2$(110) in ``constrained thermodynamic equilibrium'' with
an environment formed of O${}_2$ and CO in the complete range of 
experimentally accessible gas phase parameters. Kinetic effects due
to catalytic CO oxidation at the surface may obviously cause deviations
from this situation, and we will show in the following how the obtained
surface phase diagram may be analyzed to identify corresponding
regions in $(T, p_{\rm O_2}, p_{\rm CO})$-space where such kinetic
effects may become crucial and/or a high catalytic activity may be
expected.

In general, the surface will be close to our ``constrained
thermodynamic equilibrium'' with the reactant gas phase, as long as
the on-surface CO${}_2$ formation is the rate-limiting step. Then,
adsorption and desorption of the reactants can occur frequently on the
time scale of the reaction event, allowing the chemical potentials
of gas phase and adsorbed particles to equilibrate. In this context
we note that the only proper definition of thermodynamic equilibrium
is this equality of chemical potentials. This is not equivalent to 
saying that the adsorption and desorption rate of O and CO is equal. 
This would only be true, if no other channels for removing the particles
from the surface existed. Instead, in catalysis we can have the situation, 
that desorption {\em and} reaction compete with adsorption, but this must
not necessarily mean that the chemical potentials of the reactants 
couldn't still equilibrate with the ones in the gas phase (provided
the reaction event rate is low).

On the other hand, if the final reaction step is not rate-limiting and
the bottleneck of the reaction would e.g. be given by the adsorption 
process(es), adsorbate(s) would be faster consumed by the final reaction 
than replenished by the slow adsorption from the gas phase. Under such
conditions, the adsorbate concentration can be much lower or due to
preferential adsorption much different to what is predicted by a thermodynamic 
theory. Still, in the present case the kinetic effects will even then be 
restricted to the immediate surface population at bridge and cus sites, 
provided one stays within the range of the RuO${}_2$ bulk stability 
discussed in section IIIA, i.e. below the dotted white line in Fig. 
\ref{fig4}: the creation of vacancies in the oxide bulk lattice is
simply too costly, cf. the binding energy of O${}^{\rm 3f}$ in
Table \ref{tableII}. A Mars-van-Krevelen type mechanism, where the 
oxidation reaction is noticeably dependent on oxygen diffusion through 
the bulk \cite{mars54}, appears therefore rather unlikely for 
RuO${}_2$(110) in a more oxidizing gas phase ($\Delta \mu_{\rm O} > 
\Delta \mu_{\rm CO}$). Only if the steady-state reaction is run under 
gas phase conditions close to the dotted instability line in Fig. 
\ref{fig4} would we expect this to become different, in which case also 
oscillations between metallic and oxidic state might become 
perceivable.\cite{reuter02c}

\subsection{Catalytically most active regions}

Analyzing the surface phase diagram in Fig. \ref{fig4} along the
lines described in the preceding section, the part of the 
O${}^{\rm br}$/CO${}^{\rm cus}$ phase below the dotted instability 
line appears to be a likely region in $(T,p_{\rm O},p_{\rm CO})$-space
where catalysis might be most efficient: In contrast to all other 
phases, both reactants are then adsorbed at the oxide surface, indicating
as dominant reaction mechanism CO${}^{\rm cus}$ + O${}^{\rm br} 
\rightarrow$ CO${}_2$. We will show below that the CO${}_2$ formation 
energy barrier for this process is noticeable (1.25 eV, in good 
agreement with the result of Liu, Hu and Alavi\cite{liu01}). 
Furthermore, the energy to create vacancies in the O-bridge layer is 
rather high (2.37 eV, cf. Table \ref{tableII}), so that we would
expect the surface to be rather close to the ``constrained
equilibrium'' situation under such gas phase conditions.

The mechanism between O${}^{\rm br}$ and CO${}^{\rm cus}$ has already 
been suggested on the basis of extensive UHV experiments
\cite{over00,kim01b,fan01,wang01,wang02,kim01,seitsonen01}, where it had 
initially been characterized as a Mars-van-Krevelen type reaction, 
given that O${}^{\rm br}$ is adsorbed at the bridge sites which 
correspond to a continuation of the bulk oxide lattice stacking. As most of the 
experiments were not conducted at steady-state, but employed CO postdosage in 
UHV \cite{over00,kim01b,fan01,wang01,kim01,seitsonen01}, a further
reduction of the surface was also reported, in which after O${}^{\rm br}$
had been reacted off, CO subsequently occupied also the bridge sites
\cite{fan01,wang01,wang02,seitsonen01}. While this is fully compatible with
the higher binding energy of CO at the bridge sites, cf. Table \ref{tableII},
the phase diagram in Fig. \ref{fig4} reveals that under consideration of 
the environment, an extremely large CO/O${}_2$ partial pressure
ratio (corresponding to an almost pure CO atmosphere) would be required to
really be able to stabilize CO at the bridge sites in the 
CO${}^{\rm br}$/CO${}^{\rm cus}$ phase, cf. the given pressure scales.
Even then such a situation would not prevail for long under realistic
conditions, as RuO${}_2$ is then already instable against CO-assisted 
decomposition into Ru metal.

Hence, the initially identified part of the O${}^{\rm br}$/CO${}^{\rm cus}$
phase is still a more likely candidate for a catalytically active region.
The question remains if this is already the part in $(T,p)$-space
where we expect the RuO${}_2$ catalysis to be most efficient?
In this respect we note that a so-called stable phase in
our thermodynamic approach is not stable on a microscopic
scale, but represents an average over fluctuations of elementary processes
such as dissociation, adsorption, diffusion, association and desorption.
As all these processes and their interplay \cite{stampfl02} are of crucial
importance for catalysis, regions in $(T,\{p_i\})$-space where such 
fluctuations are particularly pronounced are expected to be most important. 
This will be the case under gas phase conditions, which correspond to
regions in the computed surface phase diagram close to boundaries between
different stable phases: At finite temperatures, the transition from one
phase to the other will not be abrupt in $(\Delta \mu_{\rm O}, \Delta \mu_{\rm CO})$-space,
but over a pressure range in which the other phase gradually becomes more
populated. The resulting phase coexistence at the catalyst surface could
then lead to a significantly enhanced dynamics, in which even additional
reaction mechanisms (in case of microscopically coexisting phases) or
reaction fronts (in case of domain pattern formation) might become operational.

Assuming a canonic distribution of the two competing phases \cite{reuter02a}
we have correspondingly estimated the region on both sides of the boundaries
in which the respective other phase is present at least at a 10\% concentration,
and marked it as white regions for the case of $T=300$\,K in Fig. \ref{fig4} 
and Fig. \ref{fig5}. Concomitantly, one of these boundary regions (the one 
between the O${}^{\rm br}$/CO${}^{\rm cus}$ and the 
O${}^{\rm br}$/O${}^{\rm cus}$ phases, henceforth termed cus-boundary),
for which we would thus expect an enhanced dynamics, falls just into gas
phase regions, for which the exceptionally high turnover rates over working
Ru catalysts have been reported, i.e. for ambient pressures in an about
equal partial pressure ratio \cite{cant78,peden86}. Along this cus-boundary
both reactants compete for adsorption at the cus sites. Provided the rather
low lateral interactions between neighboring cus sites described in section
IIIB a microscopic coexistence of both phases appears then likely, i.e. 
neighboring cus sites could be occupied independently by either O or CO. 
This would open up an additional reaction channel, CO${}^{\rm cus}$ + 
O${}^{\rm cus} \rightarrow$ CO${}_2$, which had already been observed as a 
dominant mechanism in UHV experiments on preoxidized RuO${}_2$(110), as long as
O${}^{\rm cus}$ atoms where available on these surfaces \cite{fan01,wang02}.

Below we report a rather low barrier of 0.9 eV for this process, so that
O${}^{\rm cus}$ atoms can react away rapidly under these conditions. If
the filling of empty sites with CO will also be fast as is the case in the
hatched region of the O${}^{\rm br}$/O${}^{\rm cus}$ phase close to the 
cus-boundary, then the surface will not be close to the constrained 
thermodynamic equilibrium situation and the real O${}^{\rm cus}$ concentration 
may be much lower than suggested in Figs. \ref{fig4} and \ref{fig5}. From 
our argument we therefore expect high catalytic activity in this region, but 
also note that here coverage and structure (i.e. the very dynamic behavior) 
must be modeled by statistical mechanics. Although the equilibrium approach 
thus breaks down just in this most interesting region, we see that it still 
enables us to rationalize under which gas phase parameters highest activity 
is to be expected. 

While this already helps to embed the UHV measurements into the catalytic 
context, the phase diagram also allows to systematically analyze when and 
how the pressure gap may be bridged in corresponding experiments: Namely, 
when it is assured that one stays within one phase region or along one 
particular phase boundary, assuming that then the same reaction mechanism 
will prevail. For example, at an about equal partial pressure ratio of 
O${}_2$ and CO a decrease of ambient gas phase pressures over many orders 
of magnitude will at room temperature only result in a wandering close to 
the cus-boundary in $(\Delta \mu_{\rm O}, \Delta \mu_{\rm CO})$-space, 
cf. Fig. \ref{fig4}. Correspondingly, we would conclude, that similar 
reaction rates should result, as has indeed recently been noticed by Wang 
{\em et al.} \cite{wang02} when comparing their UHV steady-state kinetic 
data with high pressure experiments of Zang and Kisch \cite{zang00}. 

However, from Fig. \ref{fig4} it is also obvious that without the knowledge
of a $(T,\{p_i\})$-phase diagram, as e.g. provided by the present work, a
naive bridging of the pressure gap by simply maintaining an arbitrary constant 
partial pressure ratio of the reactants may easily lead to crossings to other 
phase regions and in turn to uncomparable results. In this respect it is
particularly important to notice that e.g. the cus-boundary does {\em not}
exactly fall on the equal partial pressure ratio line. There is no reason
it should, and already at $T = 600$\,K the same aforedescribed decrease of
ambient gas phase pressures would bring one rapidly away from the cus-boundary
until at UHV pressures one even ends up deep inside the stability region of the
O${}^{\rm br}$/-- phase, cf. Fig. \ref{fig4}. Consequently, the computed phase
diagram leads us to predict that kinetic measurements at this temperature and
UHV pressures will result in a very low catalytic activity, as under such
conditions no CO can bind to the surface anymore. This reactivity would
then, however, be contrasted by the extremely high turnover numbers
reported for working Ru catalysts close to ambient pressures at exactly the
same temperature \cite{cant78,peden86}, in other words a typical pressure
gap situation.

\subsection{Reaction mechanisms}

In the course of the preceding section two likely oxidation mechanisms
have been identified, in which CO${}^{\rm cus}$ reacts with oxygen
adsorbed at either a neighboring bridge or a neighboring cus site.
As apparent from Fig. \ref{fig1} both channels take place along high
symmetry lines parallel to the surface: Along the $[\bar{1}10]$
direction for reaction with O${}^{\rm br}$ (dashed line) and along
the $[001]$ direction for reaction with O${}^{\rm cus}$ (dotted line),
respectively, so that we are in both cases left with a planar
configuration and may neglect the degrees of freedom perpendicular to
this plane formed by the corresponding in-surface and the out-of-surface
[110] direction. Even assuming the underlying oxide substrate as rigid, we
are thus facing at least a six-dimensional problem, involving as one possible
choice the following set of variables: Lateral and vertical position of
C${}^{\rm cus}$, lateral and vertical position of the adsorbed O, as well
as the CO bondlength and the polar angle of the CO molecular axis with respect 
to the surface. In order to shed more light on the importance of the two 
possible reaction mechanisms we proceed to determine their respective transition 
state (TS) using large $(1 \times 2)$ unit-cells to decouple the periodic images
of the reacting species. Within these cells, we map out the potential energy 
surface (PES) along two reaction coordinates, namely the lateral positions of 
the C${}^{\rm cus}$ and the adsorbed oxygen within the given reaction plane, 
while minimizing the energy at each point with respect to all remaining degrees 
of freedom (including full a relaxation of the underlying oxide lattice).

\begin{figure}
\scalebox{0.35}{\includegraphics{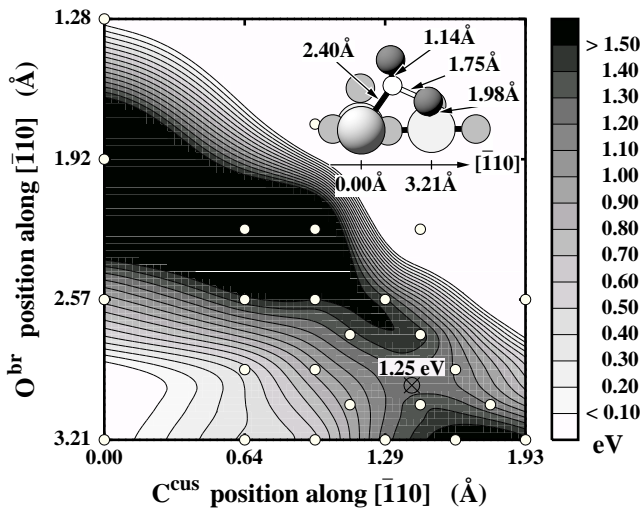}}
\caption{Potential energy surface for the reaction
CO${}^{\rm cus}$ + O${}^{\rm br} \rightarrow$ CO${}_2$.
The lateral positions of C${}^{\rm cus}$ and O${}^{\rm br}$
along the [$\bar{1}$10] direction connecting the cus site
with the neighboring bridge site, cf. the dashed line in
Fig. 1, have been constrained, fully relaxing all remaining
degrees of freedom. Actually calculated points are indicated
by white circles. The energy zero corresponds to the
initial state at (0.00 {\AA}, 3.21 {\AA}) and the transition
state geometry is shown in the inset (only the atoms lying
in the reaction plane itself are drawn as threedimensional
spheres).}
\label{fig6}
\end{figure}

Figure \ref{fig6} shows the mapped PES for the reaction CO${}^{\rm cus}$
+ O${}^{\rm br} \rightarrow$ CO${}_2$, which according to our phase
diagram is likely for a large range of gas phase parameters corresponding
to the stability region of the O${}^{\rm br}$/CO${}^{\rm cus}$ phase.
The total lateral distance between a neighboring cus and bridge site is
3.21 {\AA}, cf. Fig. \ref{fig1}, and we find the TS at a rather large
lateral C${}^{\rm cus}$ displacement of 1.44 {\AA}, while the O${}^{\rm br}$
atoms move only by 0.21 {\AA} away from their equilibrium position.
As shown in the inset of Fig. \ref{fig6} the Ru${}^{\rm cus}$-C${}^{\rm cus}$
distance is therefore remarkably lengthened at the TS (2.40 {\AA} compared to
1.99 {\AA} at the initial state), whereas the Ru${}^{\rm br}$-O${}^{\rm br}$
distance is only slightly stretched (1.98 {\AA} compared to 1.92 {\AA}). 
Looking at the binding energies compiled in Table \ref{tableII}, we can
identify as one reason for this asymmetric behavior the significantly
larger binding energy for O${}^{\rm br}$ compared to CO${}^{\rm cus}$.
Nevertheless we compute a rather low reaction barrier of 1.25\,eV
for this process, which compares favorably with the value of 1.15\,eV
reported in an earlier pseudopotential study by Liu, Hu and Alavi
employing a constrained minimization for the O${}^{\rm
br}$-CO${}^{\rm cus}$ bond length \cite{liu01}. As also pointed out in that work,
already this mechanism on RuO${}_2$(110) should have a higher reactivity
than the Ru(0001) surface, for which the CO oxidation barrier was
computed to be 1.45\,eV in a $p(2 \times 2)$ cell.\cite{stampfl97}

\begin{figure}
\scalebox{0.35}{\includegraphics{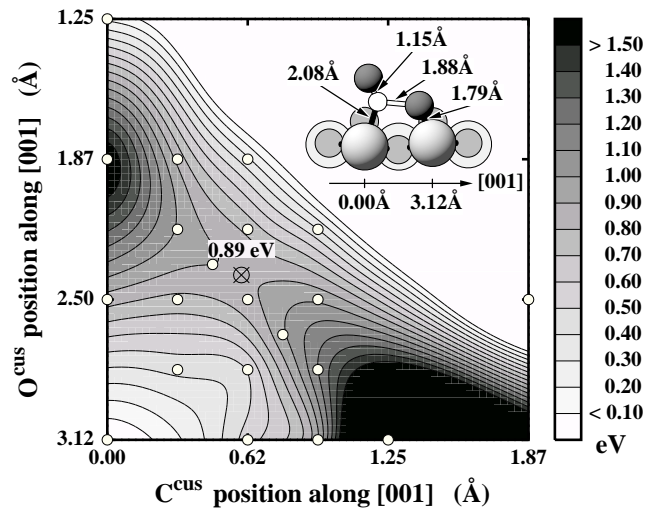}}
\caption{Potential energy surface for the reaction
CO${}^{\rm cus}$ + O${}^{\rm cus} \rightarrow$ CO${}_2$.
The lateral positions of C${}^{\rm cus}$ and O${}^{\rm cus}$
along the [001] direction connecting two neighboring
cus sites, cf. the dotted line in Fig. 1, have been 
constrained, fully relaxing all remaining degrees of freedom.
The actually calculated points are indicated by white circles.
The energy zero corresponds to the initial state at (0.00 {\AA},
3.12 {\AA}) and the transition state geometry is shown in the
inset (only the atoms lying in the reaction plane itself are
drawn as threedimensional spheres).}
\label{fig7}
\end{figure}

Turning to the second mechanism, CO${}^{\rm cus}$ + O${}^{\rm cus} 
\rightarrow$ CO${}_2$, we notice that the TS location in Fig. \ref{fig7}
has strongly shifted compared to the one displayed in Fig. \ref{fig6}.
At the TS also the O${}^{\rm cus}$ atoms have now moved laterally
by 0.73 {\AA} stretching their bondlength from 1.70 {\AA} at the
initial state to 1.79 {\AA}, so that the Ru${}^{\rm cus}$-C${}^{\rm cus}$ 
bondlength needs only to be increased to 2.08 {\AA} in comparison
to the 2.40 {\AA} required for the prior reaction. This difference
can again be understood by looking at the binding energies listed
in Table \ref{tableII}, from where the much lower bond strength of
O${}^{\rm cus}$ becomes apparent, that should facilitate an easier
displacement in comparison to the much stronger bound O${}^{\rm br}$
atoms. In line with the thus expected higher reactivity of the
O${}^{\rm cus}$ atoms, we indeed find a much lower barrier of only
0.89\,eV for this mechanism.

In view of the surface phase diagram shown in Figs. \ref{fig4} and
\ref{fig5}, this CO${}^{\rm cus}$ + O${}^{\rm cus} \rightarrow$
CO${}_2$ reaction is primarily expected under the phase coexistence
conditions along the cus-boundary. Concomitantly, it is also gas phase 
parameters that correspond to this cus-boundary, for which the
exceptionally high turnover numbers over working Ru catalysts have
been reported \cite{cant78,peden86}. Obtaining a particularly low
barrier for a reaction mechanism that may preferably be operational
under just these conditions fits thus not only nicely to the
observed high reactivity, but also underlines as specific example,
that the enhanced dynamics in phase coexistence regions might be
crucial to understand the function of surfaces under realistic
conditions.

\section{Summary}

In conclusion we have computed a {\em first-principles} based phase diagram of
the lowest energy surface structures of RuO${}_2$(110) in the ``constrained 
equilibrium'' with an environment formed of O${}_2$ and CO. Depending on the
chemical potentials of both gas phase species a number of different surface 
phases is found, ranging from two different exclusively O containing 
terminations to a completely CO covered surface. We showed how this
surface phase diagram can be analyzed to identify gas phase conditions
where kinetic effects due to the CO oxidation reaction may become crucial
and/or a high catalytic activity can be expected. In view of the possibly 
enhanced dynamics, we emphasize the particular importance of phase coexistence 
regions close to boundaries in the computed surface phase diagram, and suggest
that a reliable bridging of the pressure gap is possible, provided that one 
stays within one phase region or along one particular phase boundary. 

Concerning the specific application to the CO oxidation over RuO${}_2$(110) we 
showed that the gas phase parameters, for which exceptionally high turnover
numbers have experimentally been reported, correspond to phase coexistence 
conditions at the catalyst surface, in which the reaction CO${}^{\rm cus}$ + 
O${}^{\rm cus} \rightarrow$ CO${}_2$ may be active in addition to the hitherto 
favored CO${}^{\rm cus}$ + O${}^{\rm br} \rightarrow$ CO${}_2$. For 
the prior mechanism we calculate a significantly lower barrier of 0.89\,eV
compared to the 1.25\,eV for the latter channel, which for this specific
example underlines that phase coexistence conditions might indeed be relevant
to understand the reactivity of this catalyst surface.

Our analysis exemplifies that calculations of the kind presented in this work
can be used to identify important reaction steps at any given $(T,\{p_i\})$ in
the gas phase and to already explain a number of experimental findings on
the surface reactivity. Yet, a real microscopic description of catalysis can
only be obtained by subsequent computations of the kinetics of the manyfold of
possible elementary processes, as well as simulations addressing
the statistical interplay among them. To obtain meaningful results from the
latter type of studies the consideration of all relevant atomistic steps
is crucial, the identification of which will be greatly facilitated by
the prerequisite of knowing the various stable surface phases in equilibrium
with the given gas phase.

\section*{Appendix}

Consider a system with $N$ surface sites and a small number
of $n$ defect or adsorbate sites ($n \ll N$). Then, the
configurational entropy $S^{\rm config}$ is given by
\begin{equation}
S^{\rm config} \;=\; k_{\rm B} \; ln \frac{(N+n)!}{N! \; n!}.
\end{equation}
If we define $A_{\rm site}$ as the surface area per site, the
configurational entropy per surface area is
\begin{equation}
\frac{T S^{\rm config}}{N A_{\rm site}} \;=\; 
\frac{k_{\rm B} T}{N A_{\rm site}} \; ln \frac{(N+n)!}{N! \; n!}.
\end{equation}
For $N,n \gg 1$ we can apply the Stirling formula which gives
\begin{eqnarray}
\lefteqn{\frac{T S^{\rm config}}{N A_{\rm site}} \;=} && \nonumber \\ 
&=& \frac{k_{\rm B} T}{A_{\rm site}} 
    \left[ ln \left( 1 + \frac{n}{N} \right) +
    \left(\frac{n}{N}\right) ln \left( 1 + \frac{N}{n} \right) \right].
\end{eqnarray}
The expression in the square brackets varies between 0 (for (n/N) = 0)
and 1.2 (for (n/N)=10\%). Thus, the configurational entropy
\begin{equation}
\frac{T S^{\rm config}}{A_{\rm site}} < 1.2 \frac{k_{\rm B} T}{A_{\rm site}}.
\end{equation}

For RuO${}_2$(110), considering two sites (bridge and cus) per $(1 \times 1)$ 
unit-cell, implies that $A_{\rm site} = 10.03$\,{\AA}${}^2$. Correspondingly, 
we deduce a configurational contribution to the surface free energy of less 
than 5 meV/{\AA}${}^2$ for any $T < 1000$\,K. This is negligible everywhere 
in the phase diagram apart from the phase coexistence regions.

\end{document}